\documentclass[aps,prl,notitlepage,10pt,twocolumn,superscriptaddress]{revtex4-1}
\usepackage{bm}
\usepackage{amsmath}
\usepackage{graphicx}
\usepackage{color}

\newcommand{\ket}[1]{ |#1  \rangle}

\newcommand{\affA}{Physikalisches Institut, Universit\"at Heidelberg, Im Neuenheimer Feld 226, 69120 Heidelberg, Germany.}
\newcommand{\affB}{Institut f\"ur Theoretische Physik, Universit\"at Heidelberg, Philosophenweg 16, 69120 Heidelberg, Germany.}
\newcommand{\affC}{Max-Planck-Institut f\"ur Kernphysik, Saupfercheckweg 1, 69117 Heidelberg, Germany.}

\begin{document}

\title{Full counting statistics of laser excited Rydberg aggregates in a one-dimensional geometry}

\author{H. Schempp}\affiliation{\affA}
\author{G. G\"unter}\affiliation{\affA}
\author{M. Robert-de-Saint-Vincent}\affiliation{\affA}
\author{C. S. Hofmann}\affiliation{\affA}
\author{D. Breyel}\affiliation{\affB}
\author{A.~Komnik}\affiliation{\affB}
\author{D. W. Sch\"onleber}\affiliation{\affC}
\author{M. G\"arttner}\affiliation{\affC}
\author{J. Evers}\affiliation{\affC}
\author{S. Whitlock}\email{whitlock@physi.uni-heidelberg.de}\affiliation{\affA}
\author{M. Weidem\"uller}\email{weidemueller@uni-heidelberg.de}\thanks{also at: University of Science and Technology of China, Hefei, Anhui 230026, China}\affiliation{\affA}


\date{\today}

\begin{abstract}
We experimentally study the full counting statistics of few-body Rydberg aggregates excited from a quasi-one-dimensional atomic gas. We measure asymmetric excitation spectra and increased second and third order statistical moments of the Rydberg number distribution, from which we determine the average aggregate size. Estimating rates for different excitation processes we conclude that the aggregates grow sequentially around an initial grain. Direct comparison with numerical simulations confirms this conclusion and reveals the presence of liquid-like spatial correlations. Our findings demonstrate the importance of dephasing in strongly correlated Rydberg gases and introduce a way to study spatial correlations in interacting many-body quantum systems without imaging. 
\end{abstract}

\maketitle

Central questions in the physics of strongly-correlated many-body systems are: what is the nature of the correlations (e.g. quantum versus classical), how do they arise, and how can they be probed in real physical systems?~\cite{ichimaru1982,sachdev2001,dagotto2005,killian2007,bloch2008,carusotto2013}. Rydberg atoms with their extreme properties and long-range interactions are an ideal system to study strongly-correlated regimes, especially since the laser excitation itself in combination with strong interactions naturally produces spatial and temporal correlations~\cite{tong2004,*singer2004,reinhard2008,loew2009,amthor2010,viteau2011,schwarzkopf2011,schauss2012,viteau2012,hofmann2013}. One exciting prospect is to deterministically prepare a `quantum crystal' of Rydberg excitations by adiabatically following the ground state of the laser-dressed system~\cite{weimer2008,pohl2010,schachenmayer2010,vanbijnen2011}. However, aside from this very specific preparation scheme, the precise nature of the excitation process is not yet well understood, especially in the presence of dephasing or decoherence. An open question is whether many-body states are created simultaneously in a coherent multi-photon process or arise due to sequential excitations of individual atoms around an initial grain. Recently there has been a lot of theoretical work focusing specifically on low-dimensional systems in which longrange correlations can build up without the need for adiabatic preparation. Both resonant \cite{ates2006,ates2012,breyel2012,hoening2013,petrosyan2013a} and off-resonant \cite{robicheaux2005,weimer2010, mayle2011, lee2011, lemeshko2012, gaerttner2013} excitation have been considered, which feature different mechanisms leading to the formation of correlated structures.

\begin{figure}
\includegraphics[width=0.4\textwidth]{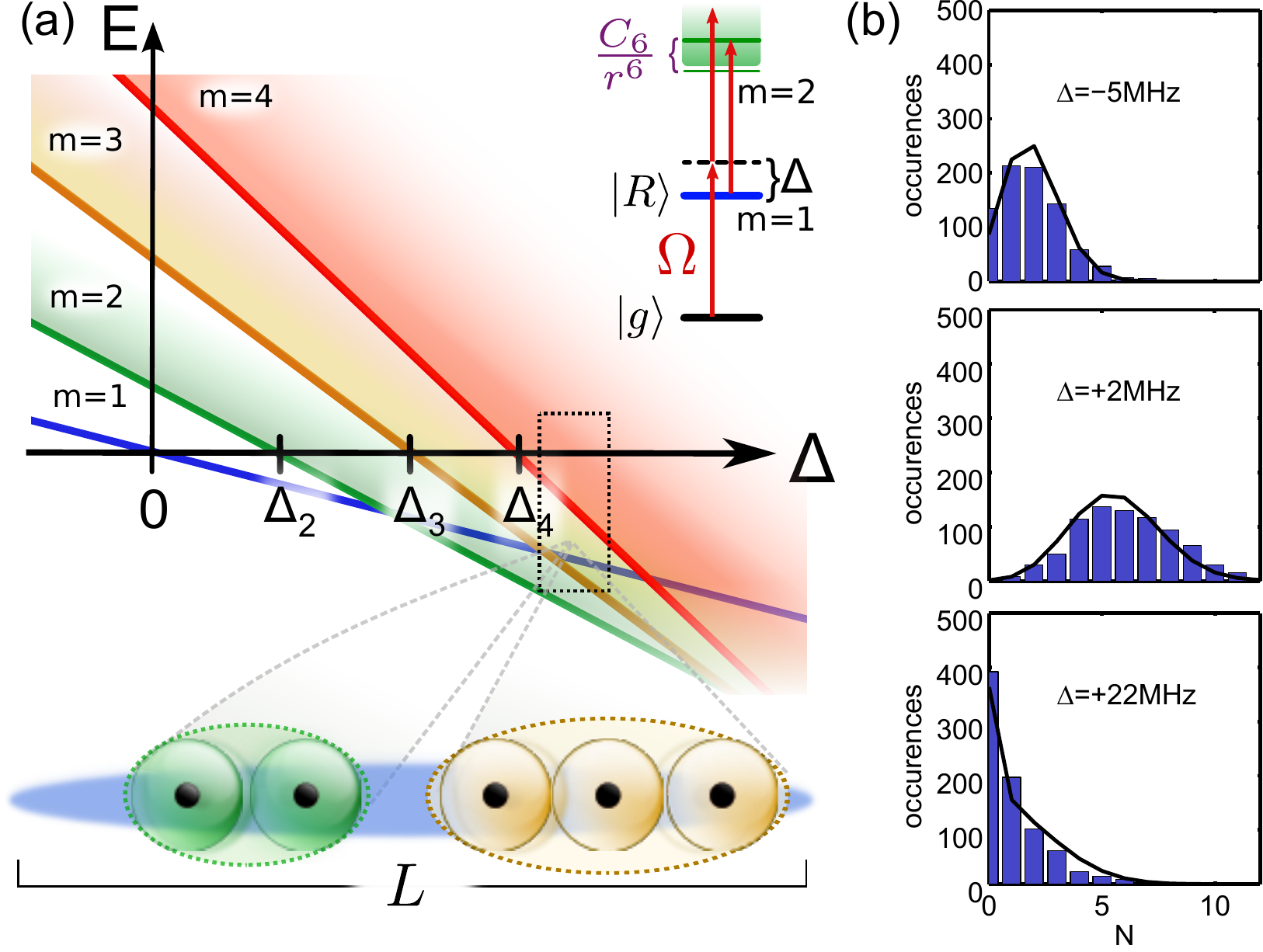}
 \caption{(Color online) (a) Level scheme and many-body energies in a rotating frame for Rydberg aggregates of size $m$ as a function of laser detuning $\Delta$. The shaded areas indicate the manifold of excited states corresponding to different spatial configurations. Zero energy crossings for the lowest energy states occur at $\Delta_m=C_6(m-1)^7/(m L^6)$ where $L$ is the system length and $C_6$ is the van der Waals interaction strength. For a given detuning and laser dephasing, aggregates of different sizes are formed (dotted rectangle), either through sequential growth or by multi-photon excitation. (b) Measured histograms of the Rydberg atom number distribution for different detunings. The solid lines are the results of the numerical simulations (see text).}
 \label{fig:fig1}
\end{figure}

Here we report the excitation of strongly correlated structures, which we call Rydberg aggregates, in a quasi-one-dimensional geometry. We make use of the full counting statistics (FCS) of the Rydberg atom number to characterize the many-body system, which serves as a complementary approach to direct imaging of spatial correlations~\cite{schwarzkopf2011,schauss2012,guenter2012}. So far, experiments on the statistics of Rydberg atoms have mainly analysed reduced number fluctuations due to the dipole blockade effect under close to resonant driving ~\cite{reinhard2008,viteau2011,viteau2012,hofmann2013}. In contrast, we interpret \textit{enhanced} number fluctuations which we observe for detuned excitation. We introduce a simplified picture to explain the effect of correlations on the FCS. We attribute the enhanced fluctuations to the excitation of Rydberg aggregates comprised of several atoms at well defined distances. To identify the dominant formation mechanisms we compare the rates for direct multi-photon excitation and sequential excitation in which an initial grain is excited slowly, followed by much faster resonant excitation at preferred distances. Our conclusions are supported by many-body simulations which include the relevant experimental parameters. 

In order to get an intuitive understanding of the excitation process we start with a simple picture for how the statistical distribution of Rydberg atoms is influenced by the laser coupling, in particular as a function of detuning. We consider Rydberg atoms in a one-dimensional geometry with repulsive interactions. Figure~\ref{fig:fig1}a shows a sketch of the bare energies of the many-body states in the rotating frame. Each energy level decreases linearly with detuning $\Delta$, with the slope proportional to the number of excitations $m$ and the offset equal to the total interaction energy~\cite{pohl2010,schachenmayer2010,amthor2010,vanbijnen2011}. The dashed rectangle represents the states that can be excited for a given detuning. 

The resonance condition for direct multi-photon excitation is $E=0$, however, aggregates can also form by an off-resonant excitation at the energy of the $m=1$ state ($E=-\Delta$) followed by additional resonant excitations. For positive detunings, configurations with positive interaction energy are energetically favored~\cite{ates2007,*ates2007a,amthor2010,gaerttner2013}. Hence we expect an enhancement of Rydberg excitations for $\Delta > 0$. Independent creation of aggregates is expected in the long-time limit of the fully coherent system \cite{gaerttner2013}. The finite duration of the excitation does not evolve the system into the global steady state, which together with the laser dephasing promotes the formation of different aggregate sizes in different parts of the excitation volume. This bunching of excitations will have a dramatic effect on higher order moments of the statistical distribution. Assuming independent excitation of nearly equally sized aggregates (same $m$), the variance of the Rydberg number $N$ scales as $\mathrm{var}(N)=m\langle N\rangle$. Higher order moments of the distribution also scale with simple powers of $m$ (see Supplemental Material~\cite{supplement}). Therefore, the resulting super-Poissonian statistics provide a measure of the typical size of aggregates. 
 
The experiments are carried out as follows. First we prepare approximately $1.5\times 10^4$ $^{87}$Rb atoms in the state $|g=5S_{1/2},F=2,m_F=2 \rangle$ in a tightly focused optical dipole trap \cite{hofmann2013a}. This results in an elongated atom cloud with e$^{-{1/2}}$ radii of $\approx 240\,\rm{\mu m}\times 1.65\,\rm{\mu m}$ (axial$\times$radial). The imaged radial cloud size of 3.5\,$\mu$m is limited by optical resolution, therefore we adjust this parameter in comparison with theory. However, it is smaller than the blockade radius which sets a limit on the closest possible distance between Rydberg atoms \cite{comparat2010}, giving rise to a quasi-1D geometry with respect to the Rydberg excitations. The maximal peak density of ground state atoms is $\approx 1.5\times 10^{12}$\,cm$^{-3}$, corresponding to a mean inter-particle spacing at the trap center of $\approx$0.9\,$\mu$m. Lower densities are achieved by reducing the time taken for initial loading 
of the dipole trap. 

Rydberg atoms in the state $\ket{R=50S_{1/2}}$ are excited by applying a two-photon laser pulse for $5~\mu$s (after turning off the optical trap). The lasers are close to two-photon resonance, detuned $\delta=$\,65\,MHz below the intermediate $|5P_{3/2},F=3,m_F=3 \rangle$ state. The first laser at $780~$nm uniformly illuminates the cloud, while the second excitation laser at $480$~nm is focused to an elliptical region of size $\approx 27\,\rm{\mu m} \times 11\,\rm{\mu m}$ (vertical $\times$ horizontal Gaussian beam waists). The two excitation laser beams counter-propagate and cross the atomic cloud perpendicular to its long axis. The effective (single-atom) two-photon Rabi frequency is $\Omega \approx 0.4$\,MHz (peak value) and the linewidth related dephasing between ground and Rydberg state is $\Gamma\approx 1$\,MHz. The two-photon detuning $\Delta$ can be varied by scanning the second step laser frequency. 

The Rydberg atoms interact repulsively with van der Waals coefficient $C_6=16$\,GHz\,$\mu$m$^6$\cite{singer2005b}. In the low density limit this gives a blockade radius of $R_c\approx 5.3 \,\rm{\mu m}$. At a density of $1.5\times 10^{12}$\,cm$^{-3}$ however we expect $N_{bl}\approx 160$ atoms per blockade sphere. As a result the Rabi frequency is collectively enhanced $\sqrt{N_{bl}}\Omega\approx 5.0$~MHz and correspondingly $R_c\approx 4.1 \,\rm{\mu m}$~\cite{loew2012}. Given our geometry we anticipate that $10-15$ Rydberg excitations are possible. After excitation we field ionize the Rydberg atoms and count the number of ions detected on a microchannel plate (MCP) detector, with an estimated detection efficiency of $\eta\approx\,$0.4~\cite{hofmann2013}. By repeating the experiment several hundred times we build up statistical distributions of the Rydberg number which are observed to have qualitatively different shapes for different detunings (Fig.~\ref{fig:fig1}b).

\begin{figure}
\includegraphics[width=0.4\textwidth]{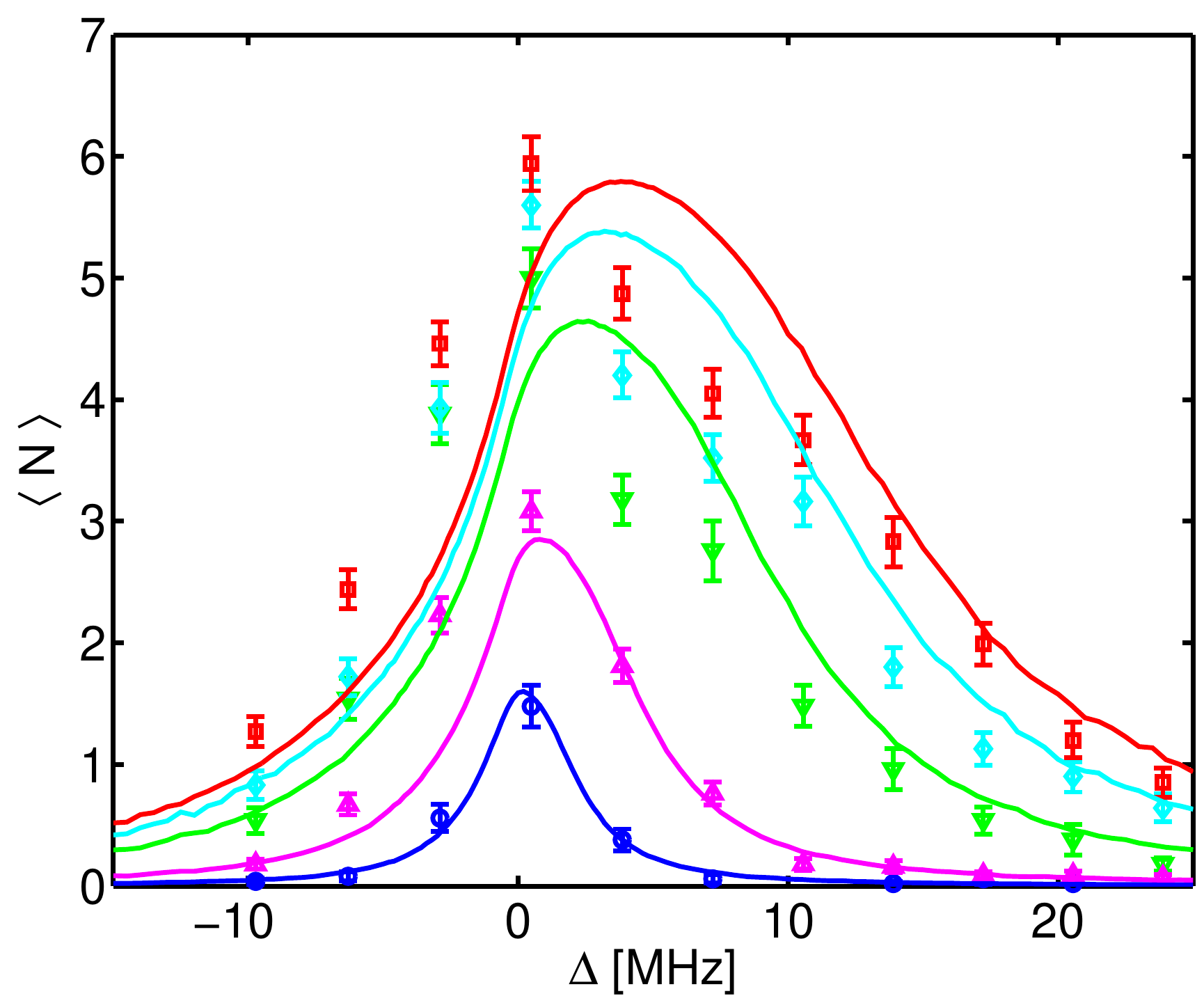}
 \caption{(Color online) Rydberg excitation spectra for different atomic densities: $5\times 10^{10}$\,cm$^{-3}$ (blue circles), $2\times 10^{11}$\,cm$^{-3}$ (cyan triangles), $8\times 10^{11}$\,cm$^{-3}$ (green triangles), $1.2\times 10^{12}$\,cm$^{-3}$ (magenta diamonds) and $1.5\times 10^{12}$\,cm$^{-3}$ (red squares). With increasing density we find enhanced excitation probabilities on the blue side of the resonance due to repulsive Rydberg-Rydberg interactions. The solid lines show the result of the rate equation model. }
 \label{fig:fig2}
\end{figure}

Figure\,\ref{fig:fig2} shows the measured mean Rydberg atom number $\langle N \rangle$ as a function of the two-photon detuning $\Delta$ for different atomic densities. At our lowest atomic densities (blue circles) the excitation spectrum is narrow and almost symmetric, reflecting the single-atom excitation probability. As the density is increased we observe a pronounced asymmetry extending to higher detunings, despite the fact that the single atom excitation probability is almost zero. This is consistent with the simple picture for the excitation of aggregates comprised of several nearby Rydberg atoms. 

The mean number of Rydberg excitations only provides partial information on the underlying many-body correlations. Extending the analysis to higher order statistical moments (full counting statistics) one can obtain additional information. This is an established technique in quantum transport problems e.g. whether electrons tunnel through a barrier as individual particles or in pairs~\cite{nazarov2009,*jehl2000,*muzykantskii1994}. To learn more about the excitation process we analyze the second-statistical moment quantified by the Mandel $Q$ parameter, defined as $Q=\langle (N-\langle N \rangle)^2 \rangle/\langle N \rangle -1$ where $N$ is the number of excitations measured in a single run of the experiment. We also analyze the third moment characterized by $Q_{3}=\langle (N-\langle N \rangle)^3 \rangle /\langle N \rangle -1$. This quantity gives an additional measure of the correlations in our system and can be related to the three-body spatial correlation function $G_3(r_1,r_2,r_3)$~\cite{ates2006}. For uncorrelated excitation of single atoms we expect $Q=Q_{3}=0$ (Poissonian limit), and assuming independent excitation of $m$-atom aggregates $Q=m-1$ and $Q_{3}=m^2-1$, respectively (see Supplemental Material~\cite{supplement}). In the presence of correlations between aggregates the resulting possible $Q$ and $Q_3$ would, depending on the degree of saturation, range from -1 to the values derived above. Estimating our statistical errors using bootstrap re-sampling~\cite{efron1979}, we conclude that measurements of fourth-order and higher moments are not statistically significant for our sample sizes. 

\begin{figure}
\includegraphics[width=0.4\textwidth]{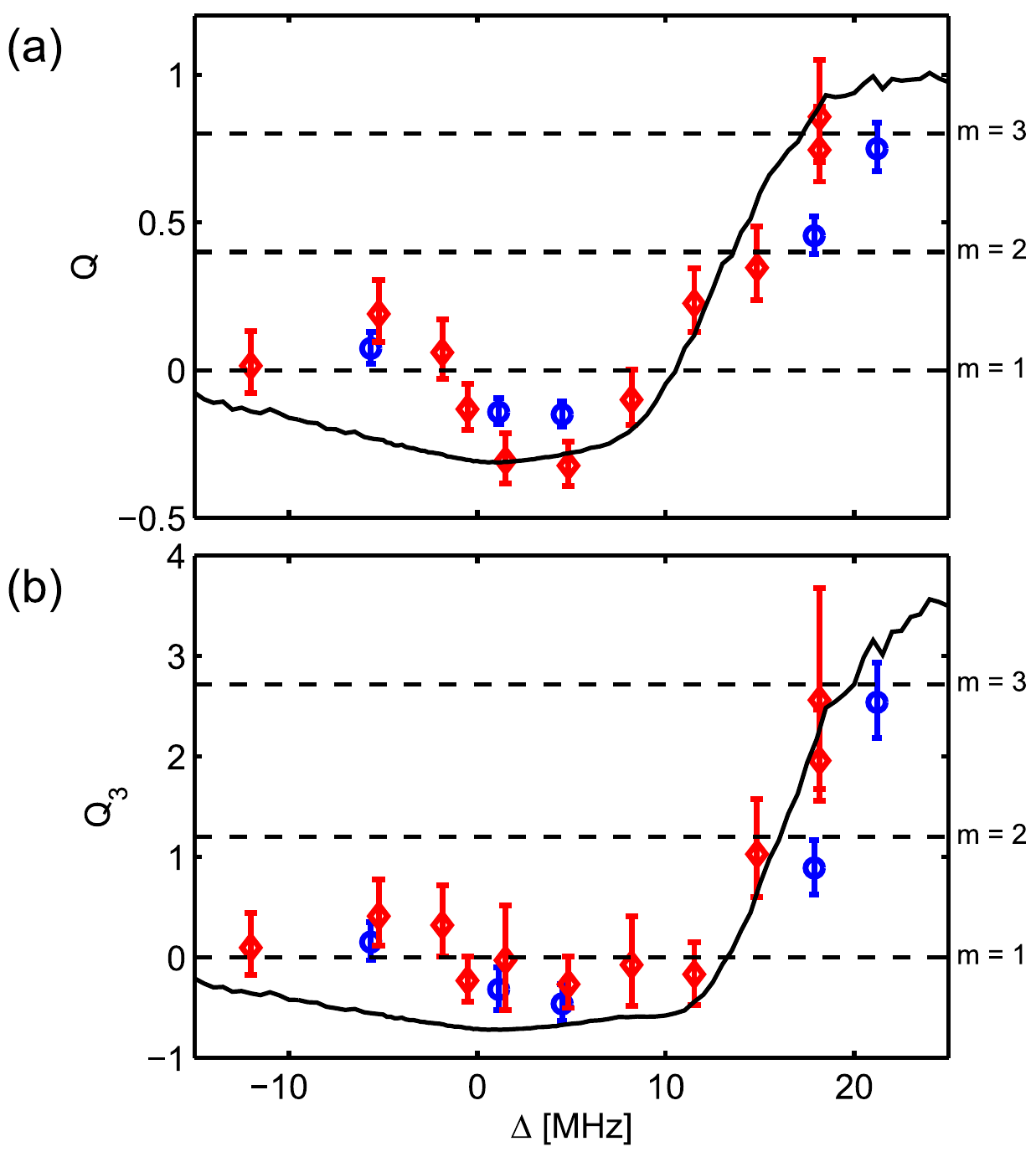}
 \caption{(Color online) $Q$ (a) and $Q_3$ (b) as a function of the detuning $\Delta$ at a density of $1.5\times10^{12} \rm{cm}^{-3}$. The red diamonds (blue circles) are extracted from a dataset with 200 (800) experiments per data point. Error bars represent 68\% confidence intervals determined via bootstrapping. The dashed lines indicate the expected $Q$ and $Q_3$ factors corresponding to the excitation of exclusively single atoms,  pairs and triples. The solid lines show the statistical moments as obtained from the rate equation model.}
 \label{fig:fig3}
\end{figure}

Figure~\ref{fig:fig3}a shows the measured $Q$ parameter as a function of detuning at the highest density of $1.5\times10^{12}~\rm{cm}^{-3}$. The red diamonds show results averaged over 200 experimental realizations while the blue circles are based on 800 measurements per point. We observe a clearly asymmetric dependence of $Q$ on the detuning. This is in marked contrast to recent observations involving Rydberg $\ket{nD}$ states featuring attractive as well as repulsive interactions, where large fluctuations were observed either side of the resonance~\cite{viteau2012}. For negative detunings we measure $Q\approx 0$ which reflects Poissonian fluctuations in the limit of weak excitation. Around resonance we find $Q$ factors clearly below 0, which indicate anti-bunching of excitations induced by the Rydberg blockade~\cite{reinhard2008,viteau2012,hofmann2013}. For $\Delta>0$ the statistical distributions become super-Poissonian ($Q>0$), which we attribute to the excitation of aggregates comprised of multiple Rydberg atoms. For $Q_3$ (Fig.\,\ref{fig:fig3}b) we observe qualitatively similar behavior to $Q$, with $Q_3\approx 0$ for $\Delta<0$, suggesting independent (Poissonian) excitation of Rydberg atoms. For $\Delta>0$ we find that $Q_3$ rapidly increases also indicating the presence of larger aggregates. 

The dashed horizontal lines in Fig.\,\ref{fig:fig3} show the simple scaling for $m$-atom aggregates, taking into account the finite detection efficiency (see Supplemental Material~\cite{supplement}). At large detunings ($\Delta\approx 20$~MHz) we find $Q\gtrsim 0.8$, consistent with an average aggregate size of $m\approx 3$ and a high probability that even larger aggregates are present. The data for $Q_3\gtrsim 2.7$ is also consistent with $m\approx 3$, thereby providing independent confirmation for the aggregate size. 
 
\begin{figure}
\includegraphics[width=0.4\textwidth]{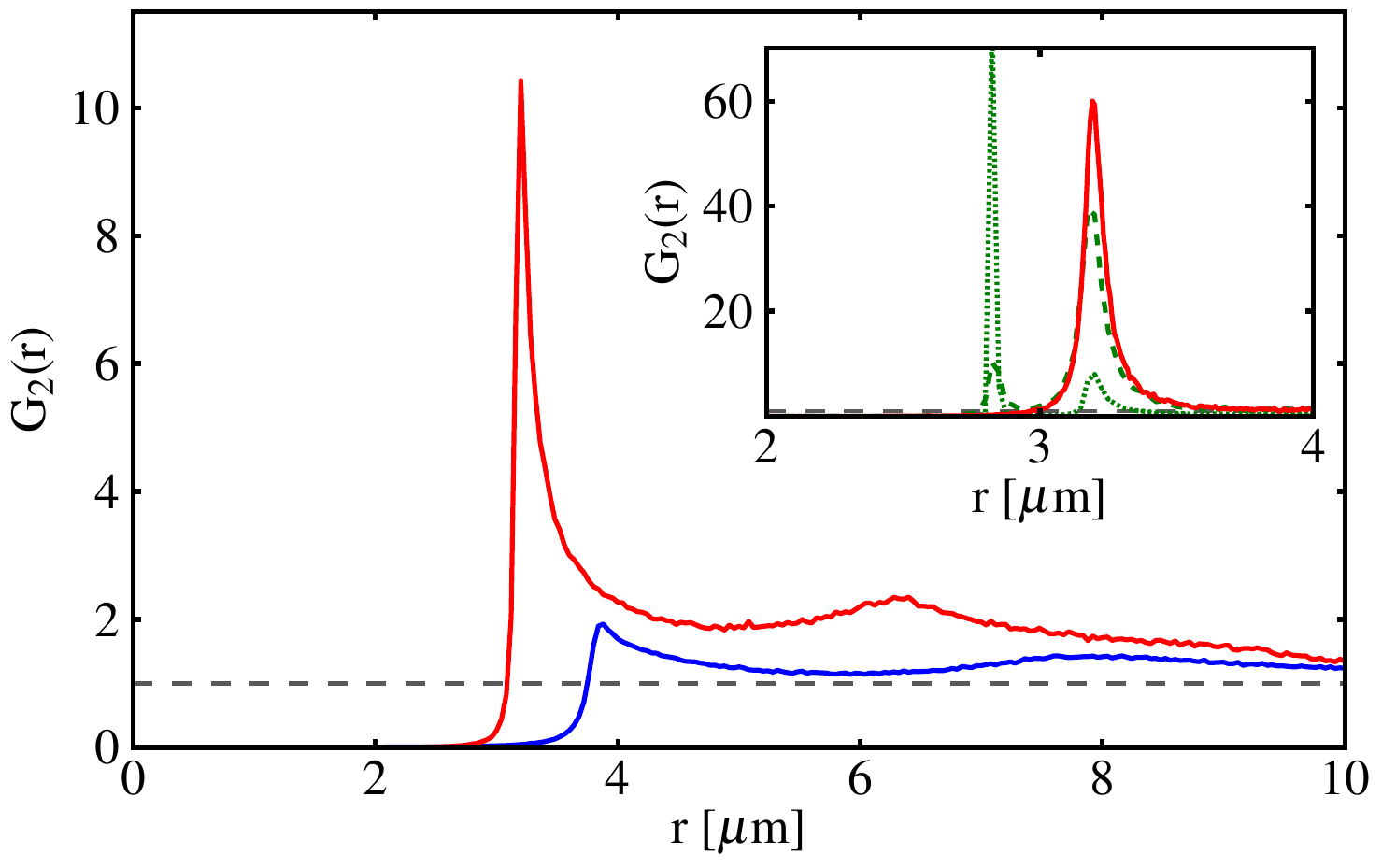}
 \caption{(Color online) Pair correlation functions $G_2(r)$ obtained from the RE model. The blue curve shows the correlation function for $\Delta=5$\,MHz, the red curve for $\Delta=15$\,MHz. 
The inset shows MCWF simulations for $\Delta=15$\,MHz, and dephasing rates $\Gamma=$0 (green, dotted) and $\Gamma=1$\,MHz (green, dashed), compared to the RE simulation with $\Gamma=1$\,MHz (red, solid). To improve visibility, the dotted curve is scaled by a factor of $1/10$. The different peak amplitudes between the inset and main figure are due to the different simulation volumes and finite size effects.}
 \label{fig:fig4}
\end{figure}

To address the question of how the aggregates form we estimate the dominant formation rates in the limit of large laser detuning. The overall rate for sequential excitation is limited by the off-resonant excitation rate of the first atom, $\gamma_1\approx\Omega^2\Gamma/(4\Delta^2)$, since subsequent excitation steps are resonant for the preferred distance $r=(C_6/\Delta)^{1/6}$. In comparison, if the total interaction energy for a state involving $m$ excitations is precisely matched by the laser detuning, then simultaneous $m$-photon excitation can occur with a rate $\gamma_{m,sim}=(\Omega^m/(2^{m-1}\prod_{i=1}^{m-1}{\delta_i}))^2/(m \Gamma)$, obtained by adiabatically eliminating all intermediate states. The detunings from the intermediate states involving $i<m$ excitations (assumed to be equidistantly spaced along a line) are $\delta_i=i \Delta -(i-1) m \Delta /(m-1)$. The ratio of sequential to simultaneous rates for $m$-atom aggregates therefore scales as $\Gamma^2 \Delta^{2 m-4}\Omega ^{2-2 m}$. Accounting for the availability of atoms at specific distances leads to a slight modification of these rates, nonetheless, for $\Omega < \Delta$ the ratio still increases exponentially with aggregate size $m$. For our experimental parameters the rate for simultaneous $m=2$-photon excitation is approximately one order of magnitude smaller than sequential growth. For $m=3$ simultaneous excitation is suppressed by an additional factor of $\approx4\times 10^3$ at $\Delta$=20\,MHz.

To further elucidate whether sequential excitation is the dominant formation mechanism in our experiments we perform time-dependent numerical simulations using an effective two-level rate equation (RE) model (\cite{heeg2012}, based on~\cite{ates2007,*ates2007a}). The RE approach can model our precise experimental geometry, but approximates many-body correlations via an energy shift depending on the state of the surrounding atoms. Thus it does not capture multi-atom coherences which would be present in a simultaneous multi-photon excitation. Nonetheless, we find that the results of the RE model qualitatively reproduce the full statistical distributions including higher order moments for $\Delta>0$ where the aggregates are formed (solid lines in Figs. 1b, 2 and 3). We observe a small discrepancy between theory and data on the red side of the resonance ($\Delta<0$) and a slight shoulder in the mean number around $\Delta\approx$10\,MHz which cannot be explained by the RE model. This shoulder possibly indicates the presence of additional physical processes like atomic motion due to repulsive forces which would be more pronounced on the blue side of the resonance because of the excitation of Rydberg atoms with small separations. Nevertheless, given the good agreement, especially for the Mandel Q parameter in the regime $\Delta>0$, we conclude that coherent multi-photon excitation is not required to explain our experimental findings, since such processes are not included in the RE treatment. Instead, for our parameters the dominant mechanism by which aggregates are formed at large detuning is via sequential (single-atom) excitations around an initial grain. This is further supported by time-dependent simulations which show that the off-resonant excitation of the initial grain is slow while the subsequent resonant excitation of additional atoms happens on faster time scales, as we show in the Supplemental Material~\cite{supplement}. 

To further substantiate that sequential excitation dominates over direct multi-photon excitation we benchmark the RE model with two-level wave function Monte-Carlo (MCWF) simulations. The MCWF method includes multi-atom coherences, but can only be applied to systems involving fewer excited atoms. Therefore, we simulate 50 atoms in a spherical volume with radius $2.285~\mu$m and the other parameters being comparable to those of the experiments. The close agreement of the two simulations (see Fig.\,1 in the Supplemental Material~\cite{supplement}), together with the good agreement between RE model and experimental data is a strong indication for the sequential excitation indeed being the dominant process in our system.

Considering the good agreement between theory and experimental data, we can extract more information about the underlying many-body correlations from the results of the RE model. Specifically, we extract the second order spatial correlation function $G_2 (r)$ (see Fig.~\ref{fig:fig4}), as defined in Ref.~\cite{heeg2012}. For $\Delta=+15$\,MHz we observe strong liquid-like correlations (i.e. a strong first peak in $G_2 (r)$ followed by peaks with decaying amplitudes) which are responsible for the large measured $Q$ values. The pronounced peak at $r = (C_6 /\Delta)^{1/6}\approx 3.2~\mu$m reflects a strongly preferred pair distance. A smaller peak at $r \approx 6.5 \,\rm{\mu m}$ is evidence for higher order correlations for $m > 2$. The peaks are strongly reduced for near resonant excitation ($\Delta =+5$\, MHz). For resonant driving the peaks are even smaller, but still non-trivial correlations are present, as recently demonstrated experimentally \cite{schauss2012}. These spatial correlations could be exploited in other areas of physics such as in the creation of strongly-coupled plasmas~\cite{killian2007,robert-de-saint-vincent2013,bannasch2013}. Comparing to the $G_2 (r)$ as obtained from MCWF simulations (100 atoms in a cylindrical volume with radius $1.65~\mu$m and length $6~\mu$m) we find again close agreement (Fig.~\ref{fig:fig4} inset), apart from a small additional peak at $r = (C_6 /2\Delta)^{1/6} \approx 2.8~\mu$m, which indicates a minor contribution from direct excitation of Rydberg atom pairs. This peak depends sensitively on the laser dephasing which quickly destroys multi-atom coherences. 


In conclusion we have investigated laser excited Rydberg aggregates in a quasi-1D geometry. Using full counting statistics we determine their typical size as $m\gtrsim 3$. Our work emphasizes sequential excitation as the dominant mechanism under conditions of large dephasing. This highlights the need to account for dephasing and dissipation in theoretical descriptions of strongly correlated Rydberg gases. So far, most theoretical work has focused on lattice geometries~\cite{mayle2011,lee2011,lemeshko2012,ates2012,hoening2013}. It remains an open question whether results  qualitatively different from ours are obtained for off-resonant excitation in optical lattices. However, even for lower dephasing rates, the vanishing multi-photon Rabi frequencies for large $m$ suggests that sequential growth of aggregates is likely to play a significant role in existing experiments~\cite{schauss2012}. Distinguishing sequential and simultaneous excitation using FCS is intrinsically challenging, since both processes lead to very similar results. Hence, in order to experimentally distinguish coherent multi-photon excitation vs. sequential growth, future experiments could e.g. measure the double peak structure in the spatial correlation function $G_2(r)$ as shown in Fig.\,4 (inset). 

We thank T. Pohl, O. Morsch, I. Bloch, M. H{\"o}ning and M. Fleischhauer for fruitful discussions. This work is supported in part by the Heidelberg Center for Quantum Dynamics, the Deutsche Forschungsgemeinschaft under WE2661/10.2 and the EU Initial Training Network COHERENCE. M.R.D.S.V. acknowledges support from the EU Marie-Curie program (Grant No. FP7-PEOPLE-2011-IEF-300870).

\clearpage

\section{Supplemental Material}

\subsection{Statistical distributions for aggregate excitation}
The $p$-th central moment of a distribution $P(K)$ is defined as
\begin{equation}
\mu_{K,p} = \langle(K-\langle K \rangle)^p \rangle
\end{equation}
We consider the excitation of aggregates containing a fixed number of atoms $m$. The number of aggregates is $K$, such that the number of Rydberg excitations is $N=mK$. For an aggregate number distribution $P(K)$, we find for the $p$-th central moment of the excitation number distribution
\begin{equation}
\mu_{N,p} = \langle(m K-\langle m K \rangle)^p \rangle = m^p \langle(K-\langle K \rangle)^p \rangle.
\end{equation}
Each central moment is enhanced by a factor $m^p$ as compared to the respective central moment of the distribution of aggregate numbers $K$. Assuming the aggregates are excited independently (which is justified considering the low excitation numbers at large detunings indicating the system is far from saturation) $P(K)$ is a Poissonian distribution, which means that the second and third central moment equal the mean of the distribution. With the mean atom number $\langle N \rangle = m \langle K \rangle$ we find
\begin{align}\label{eq:qvsm}
Q & = \frac{\langle (N-\langle N \rangle)^2 \rangle}{\langle N \rangle}-1 = m-1 \nonumber \\
Q_3 & =  \frac{\langle (N-\langle N \rangle)^3 \rangle}{\langle N \rangle}-1 = m^2-1
\end{align}

\subsection{Effect of finite detection efficiency on statistics}
The detection efficiency $\eta\approx 0.4$ in our experiment affects the measured statistical distributions and it must be taken into account for a valid comparison with theory. We assume the detection process to be independent for each Rydberg atom and also to not be influenced by the total number of Rydberg atoms, which excludes saturation effects. In this case we can model the detector operation as a binomial process. The probability distribution of physically present Rydberg atoms $P(N)$ and the probability distribution of the detected Rydberg atoms $P^\prime(N^\prime)$ are related by
\begin{equation}
 P^\prime (N^\prime) = \sum\limits_{N=0}^{\infty} f_\eta(N^\prime | N) P (N),
 \label{eq:transformation}
\end{equation}
where for all practical purposes the upper limit of the sum may be set to a finite value and the kernel $f_\eta(N^\prime | N)$ of this transformation is given by
\begin{equation}
 f_\eta(N^\prime | N) = \left( \begin{array}{c} N \\ N^\prime \end{array} \right) \eta^{N^\prime} (1-\eta)^{N-N^\prime}.
 \label{eq:binomial}
\end{equation}
Assuming $\eta$ is known, equation \eqref{eq:transformation} now allows to relate the $p$-th order physical moments $\langle N^p \rangle = \sum N^p P(N)$ to the measured moments $\langle {N^\prime}^p \rangle = \sum {N^\prime}^p P^\prime(N^\prime)$ of their respective distributions. The results read 
\begin{align}
 \langle N^\prime \rangle & = \eta \langle N \rangle \notag \\
 \langle {N^\prime}^2 \rangle & = \eta^2 \left( \langle N^2 \rangle - \langle N \rangle \right) + \eta \langle N \rangle \quad \text{and} \notag \\
 \langle {N^\prime}^3 \rangle & = \langle N \rangle (\eta - 3 \eta^2 + 2 \eta^3) + \langle N^2 \rangle (3 \eta^2 - 3 \eta^3) + \eta^3 \langle N^3 \rangle,
\end{align}
and can be used to calculate the corresponding cumulants $\langle \! \langle {N^{\prime}}^p \rangle \! \rangle$ \cite{gut2010}. Subsequently the measured Mandel-$Q$-parameter, generalized to $i$-th order,
\begin{equation}
 Q_i^{\prime} = \frac{\langle \! \langle {N^{\prime}}^i \rangle \! \rangle }{\langle N^{\prime} \rangle} - 1.
\end{equation}
can be related to the $Q_i$ of the true distribution. Evaluating the above for variance $\langle \! \langle {N^{\prime}}^2 \rangle \! \rangle $ and third cumulant  $\langle \! \langle {N^{\prime}}^3 \rangle \! \rangle $ we get

\begin{align}
 Q_2^{\prime} & = \eta Q_2 \quad \text{and} \notag \\
 Q_3^{\prime} & = 3 Q_2 (\eta - \eta^2) + Q_3\eta^2.
\end{align}
where $Q_2:=Q$ in the main text and eq.\eqref{eq:qvsm}.

\subsection{Benchmark of the RE model with MCWF simulation}

To benchmark the RE simulation, we compare its results with simulation results from a two-level wave function Monte-Carlo (MCWF) simulation. The MCWF simulation is essentially exact (including the full many-body correlations), but can only be applied to smaller systems involving fewer excited atoms. Accordingly, we consider for the comparison a spherical trap with radius $r=2.285\, \rm{\mu m}$ and density $1\times 10^{12}$ cm$^{-3}$. Effective laser parameters are $\Omega = 0.3\,\mathrm{MHz}$, $C_6 = 16\,\mathrm{GHz} \, \rm{\mu m}^6$, $\Gamma = 1\,\mathrm{MHz}$ and the population decay from the Rydberg state $\gamma = 0.004\,\mathrm{MHz}$. The dynamics is evaluated after $t=5\,\rm{\mu s}$.

\begin{figure}[t!]
\centering
\includegraphics[width=\columnwidth]{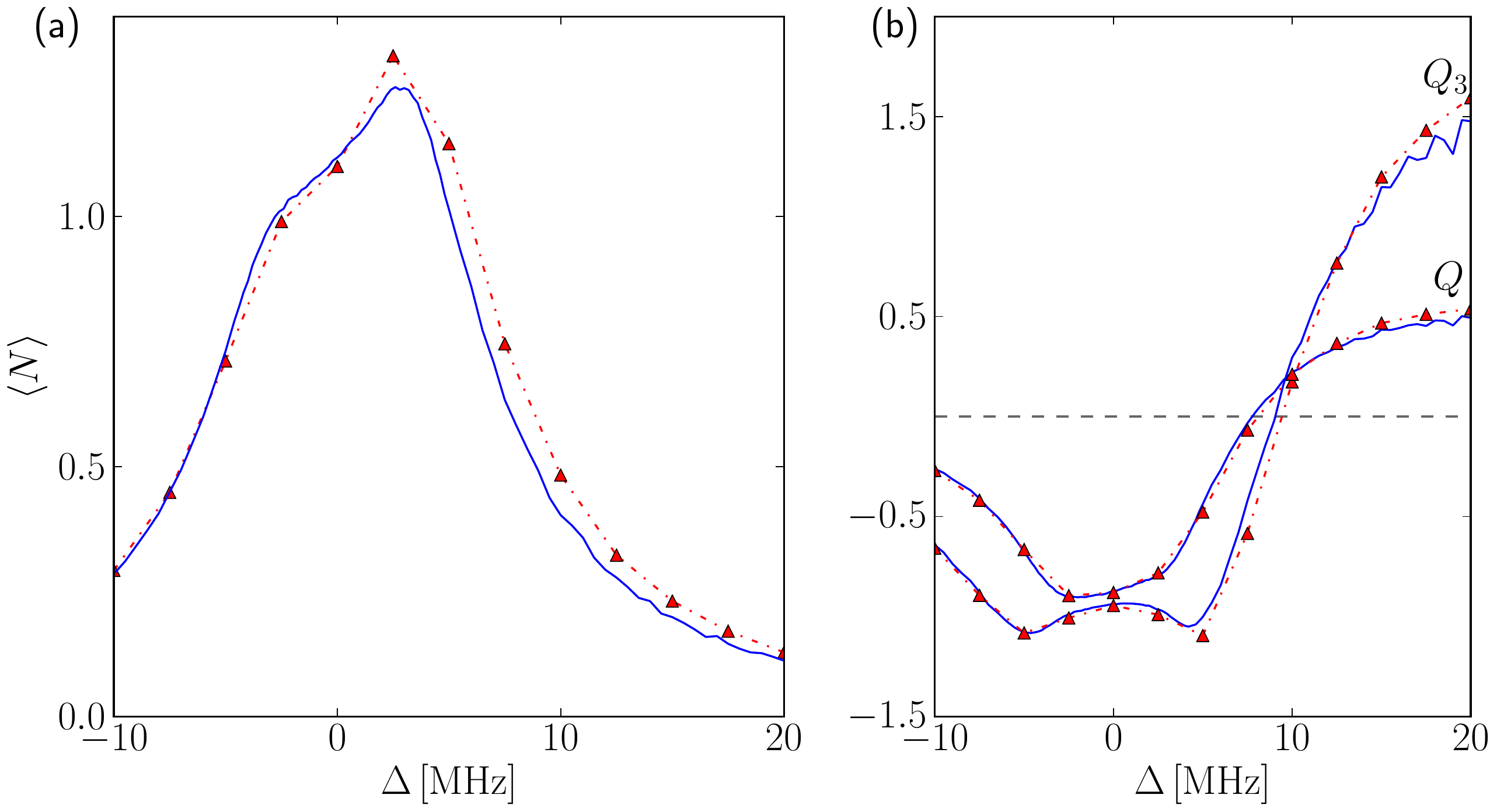}
\caption{\label{fig:MCWF_compare}(Color online) Rydberg population (a) and generalized Mandel parameters $Q$ and $Q_3$ (b) as a function of the laser detuning. Rate equation (blue, solid line) and MCWF simulation (red triangles) for parameters given in the supplement text.}
\end{figure}

\begin{figure}[t!]
\centering
\includegraphics[width=8cm]{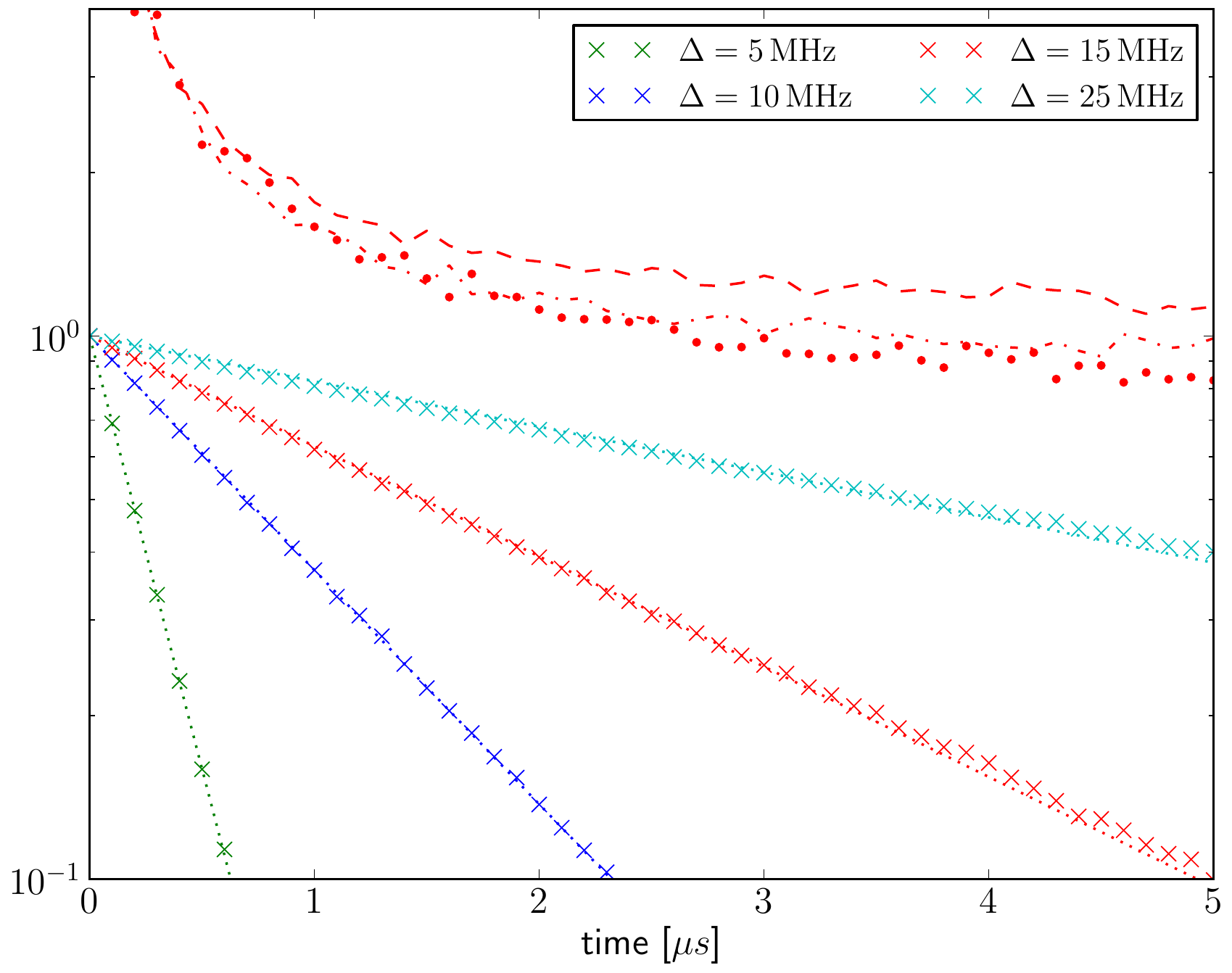}
\caption{\label{fig:origin}(Color online) Time evolution of the populations of the different excitation number subspaces $P(N)$ with $N$ excitations. The four curves (crosses) in the lower part of the figure show the time evolution of the zero excitation subspace $P(0)$ for different detunings of the incident laser field $\Delta = 5,10,15,25$~MHz, from bottom to top. The data points obtained from the RE model are overlaid with an exponential fit. 
The three curves (dashed, dash-dotted, dotted) in the upper half of the figure show the time evolution of ratios $P(N)/P(N+1)$ of excitation probabilities of adjacent subspaces for $N \in \{1,2,3\}$. From top to bottom, ratios with increasing $N$ are shown.}
\end{figure}

Figure~\ref{fig:MCWF_compare}a shows the Rydberg population as a function of the laser detuning. Both the RE simulation (blue, solid line) and the MCWF simulation (red triangles) closely agree. As this also holds for the higher order generalized Mandel parameters $Q$ and $Q_3$ (cf. Fig.~\ref{fig:MCWF_compare} b), we conclude that the rate equation model captures essentially all features of the experimental data and can be used to model the experiment. 
\\

\subsection{Aggregate formation mechanism}

Here we elaborate on the relation between the aggregate formation mechanism and the observed statistics. We analyze the time evolutions of the individual probabilities $P(N)$ for $N$-fold Rydberg excitation  of the ensemble, as produced by the RE simulations. In Fig.~\ref{fig:origin}, we show $P(0)$ as a function of time for different detunings $\Delta = \{5, 10, 15, 25\}$~MHz on a logarithmic scale. In all cases, the population in the zero-excitation state decreases exponentially until a significant part of the population is excited to higher $N$ values. Furthermore, the depopulation of the zero-excitation state becomes slower with increasing detuning. Fitting the exponential decay rates, we find that the obtained rates approximately scale with $1/\Delta^2$, in particular for higher detunings. This is compatible with the expectation for a detuning-dominated rate equation modeling. At intermediate times comparable to the exposure time in the experiment, we observe evidence for bimodal distributions.

Next, the time evolution of the higher excited subspaces $P(N>0)$ is analyzed with fixed detuning $\Delta = 15$~MHz. Rather than studying them individually, Fig.~\ref{fig:origin} shows ratios $P(N)/P(N+1)$ for $N \in \{1,2,3\}$. Initially, these ratios are large, and they rapidly converge to an approximate stationary state $\gtrsim 1$ on a time scale much faster than the decay out of the zero-excitation subspace $P(0)$. 
The reason for this is that only the first excitation is off-resonant, while subsequently the aggregate size equilibrates rapidly by resonant excitation of atoms with well defined interaction energy. The slow timescale for the formation of the initial seed excitation and rapid subsequent growth leads to enhanced excitation number fluctuations, and thus an increased value for the Mandel $Q$ parameter.

\subsection{Aggregate formation rates}
The timescale for the sequential formation of an aggregate is the sum of the timescales of the individual excitation steps. Therefore it is dominated by the slowest process, which is the off-resonant excitation of the initial grain.  The rate for this process is given by $\gamma_1\approx \Omega^2 \Gamma/(1+4 \Delta^2/\Gamma^2)$ (see e.g. \cite{grynberg1977}) which simplifies to $\Omega^2 \Gamma/(4 \Delta^2)$ if $\Delta\gg\Gamma$. 

The rate for an overall resonant simultaneous $m$-photon excitation can be obtained by adiabatically eliminating the intermediate states, i.e. the states containing between 1 and m-1 excitations. For a two-photon excitation, e.g., the rate is given by $\gamma_{2,sim}=(\Omega^2/(2 \delta))^2/\Gamma$ with $\delta$ being the detuning from the intermediate state (see e.g. \cite{grynberg1977}). For m-photon excitation the formula can be generalized to $\gamma_{m,sim}=(\Omega^m/(2^{m-1}\prod_{i=1}^{m-1}\delta_i))^2/(m \Gamma)$ where $\delta_i$ is the detuning from the intermediate state containing $i<m$ excitations. The adiabatic elimination is valid as long as all intermediate detunings $\delta_i$ are smaller than both the dephasing rate and the respective multi-photon Rabi frequency. For our experimental parameters these criteria are fulfilled for aggregates of size up to $m=6$ assuming approximately equidistant spacing between the atoms. This line of reasoning holds for densities high enough that atoms are available for excitation at the required distance from the atoms excited in the first place.

%

\end{document}